# Spatial confinement, magnetic localization and their interactions on massless dirac fermions

Zhong-Qiu Fu[§], Yu Zhang[§], Jia-Bin Qiao, Dong-Lin Ma, Haiwen Liu, Zi-Han Guo, Yi-Cong Wei, Jing-Yi Hu, Qian Xiao, Xin-Rui Mao, and Lin He[*]
Center for Advanced Quantum Studies, Department of Physics, Beijing Normal University, Beijing, 100875, People's Republic of China

[§]These authors contributed equally to this work.

[*]Correspondence and requests for materials should be addressed to L.H.
(e-mail: helin@bnu.edu.cn ).

**It is of keen interest to researchers understanding different approaches to confine massless Dirac fermions in graphene, which is also a central problem in making electronic devices based on graphene. Here, we studied spatial confinement, magnetic localization and their interactions on massless Dirac fermions in an angled graphene wedge formed by two linear graphene *p-n* boundaries with an angle ~ 34 °. Using scanning tunneling microscopy, we visualized quasibound states temporarily confined in the studied graphene wedge. Large perpendicular magnetic fields condensed the massless Dirac fermions in the graphene wedge into Landau levels (LLs). The spatial confinement of the wedge affects the Landau quantization, which enables us to experimentally measure the spatial extent of the wave functions of the LLs. The magnetic fields induce a sudden and large increase in energy of the quasibound states because of a π Berry phase jump of the massless Dirac fermions in graphene. Such a behavior is the hallmark of the "Klein tunneling" in graphene. Our experiment demonstrated that the angled wedge is a unique system with the critical magnetic fields for the π Berry phase jump depending on distance from summit of the wedge.**

**INTRODUCTION**

Understanding microscopic mechanisms of different approaches in confinement of massless Dirac fermions is a central problem in fields ranging from condensed matter to relativistic quantum mechanics. In graphene, the low-energy charge carriers behave as the massless Dirac fermions, providing an ideal platform to explore this subject experimentally (*1-16*). For example, the massless Dirac fermions in graphene monolayer can be temporarily trapped to form quasibound states by spatial-varying electrostatic potentials (*6,7,11,12*) or completely localized into Landau levels (LLs) in large perpendicular magnetic fields (*17-22*). By combining two different methods, i.e., the electrostatic potentials and the magnetic fields, in the confinement of the massless Dirac fermions, it is interesting to note that the studied graphene system could exhibit exotic properties that are beyond those already imaged (*2,8,9,13,16*). For example, the π Berry phase of quasiparticles in a graphene quantum dot can be switched on and off by the magnetic fields (*2,13*), which cannot be realized by purely using electrostatic potentials or magnetic fields (*23-25*).

In this paper, we carried out measurements in an angled graphene wedge formed by two linear graphene *p-n* boundaries with an angle ~ 34°. Effects of magnetic fields on the properties of quasibound states confined between two parallel *p-n* boundaries of graphene was first studied theoretically (*23*). It was predicted that magnetic fields will result half-period shift in the Fabry-Pérot fringe pattern. Such a phenomenon can be treated as a hallmark of Klein physics in graphene and was demonstrated explicitly in experiment by transport measurements across two linear and parallel *p-n* junctions (*2*). Here, we studied spatial confinement, magnetic localization, and their interactions on the massless Dirac fermions in the angled graphene wedge by using scanning tunneling microscopy and spectroscopy (STM and STS). Our experiment indicates that the spatial confinement affects the Landau quantization in the graphene wedge, and the magnetic fields also strongly influence the confinement induced quasibound states by sudden jump in binding energy at certain critical magnetic field $B_C$, which originates from the π Berry phase shift of the closed trajectories. Interestingly, the critical magnetic fields for the π Berry phase jump depend on distance from summit of the wedge.

**RESULTS**

In our experiment, graphene was produced by low pressure chemical vapor deposition on a copper pocket (*26*) (see Method and Fig. S1 for details of growth (*27*)) and about half coverage on outside surface of the Cu pocket was bilayer graphene (see Fig. S2 (*27*)). Figure 1A shows a typical three-dimensional STM topographic image of the angled graphene wedge. Two clear steps are observed in the height profile line (Fig. 1B). The smaller step (~110 pm) may be caused by an intercalation of $Cu_xO$ layer according to a previous study (*28*). The larger step (~320 pm), which is the left boundary of the wedge, is generated by an underlying graphene layer. The right boundary of the wedge is along a sharp step edge of the substrate. Zoom-in STM image of white dashed square in Fig. 1A and its corresponding fast Fourier transform (FFT) pattern are shown in Fig. 1C. Besides the atomic resolution of graphene, moiré pattern like structure is observed. Correspondingly, the FFT pattern shows two other periodic superlattice peaks except the Bragg peak of graphene (six black circles). The four rhombic vertices marked by red circles correspond to the sulfur nanocluster superlattices (*29*) (the substrate is S-rich Cu foil and the S atoms segregate from the Cu foil during the growth process, see Fig. S3 for further discussion (*27*)), and the six hexagonal points marked by yellow circles are generated by the moiré pattern (~1.09 nm) of two adjacent graphene layers with a twisted angle about 6.5°. Such a large rotational misalignment usually leads to the decoupling of the two graphene sheets (*30-33*). Figure 1D shows the schematic structure of the studied system.

Though the topmost graphene sheet is continuous in the studied region, the local work function on and off the angled wedge is different. Field emission resonances (FER) were carried out to measure their differences. According to the energy shifts of the first FER peak, the differences of the local work function on and off the wedge are about 80 meV on left boundary and are about 60 meV on right boundary (see Fig. S4 for the details (*27*)). The Dirac point $E_D$ in the angled wedge is determined as about -10 meV according to the zero LL of high-field STS spectra (see Fig. S5 (*27*)). Therefore, the $E_D$ out the angled wedge should be at ~ 70 meV (in the left region) and ~ 50 meV (in the right region). Our experiment indicates that the graphene wedge is formed by two linear *p*-*n* junctions with an angle of about 34 °, as schematically shown in Fig. 2A. Within the angled wedge, we can clearly observe interference patterns of the massless Dirac fermions due to the formation of the *p*-*n*-*p* junction (see a movie for STS maps taken from 0 to 400 mV (*27*)). Figure 2B shows a representative STS map recorded at 167 mV, which exhibits interference pattern originating from Klein scattering at the *p*-*n* boundaries, i.e., there is perfect transmission for normal incidence and a finite reflection at non-normal incidence (*34*). The average wavelength $\lambda$ of the interference pattern at 167 meV is determined as about 24 nm. In our experiment, the $\lambda$ decreases with the increasing energy *E*, as shown in Fig. S6 (27) and their relation can be described quite well by the expression $\lambda = \frac{hv_F}{E}$ ($h$ is Planck constant, $v_F \approx 10^6$ m/s is the Fermi velocity), demonstrating explicitly the trapping of the massless Dirac fermions in the angled wedge. For a continuous graphene sheet on Cu steps, there is no signal of spatial confinement of Dirac ferimions (see Fig. S7) because there is no difference of local work function on and off the step, and the scattering of the boundary of Cu steps is negligible.

The confinement of the massless Dirac fermions in the angled wedge is further probed by measuring the tunneling differential conductance, $g(V_b, l, B) = dI/dV_b$, as a function of distance from the measured position to the summit of the wedge, *l*, and magnetic field, *B*. Figure 2C shows the tunneling differential conductance versus $l$ obtained along the black arrow in Fig. 2A in zero magnetic field. A series of tunneling resonances are clearly observed, especially above the Dirac point of the *n*-doped region, as shown in Fig. 2C and the inset of Fig. 2D. Such a result is quite reasonable. For a *p*-*n*-*p* junction, massless Dirac fermions above the Dirac point of the *n*-doped region are expected to be temporarily trapped to form the quasibound states. According to our experimental result, the average energy spacing $\Delta E$ between the quasibound states decreases linear with increasing *l*, as shown in Fig. 2D. This agrees quite well with the expected level spacing of the quasibound states confined in a region with the spatial size $l$, $\Delta E = \pi\hbar v_F/l$ (here $\hbar$ is Planck's constant divided by $2\pi$) (*6,7*).

In the graphene wedge, the topmost graphene sheet electronically decouples from the supporting substrate and we can observe well-defined LLs in large perpendicular magnetic fields (Fig. S6 (*27*)). Figures 3A-3D show differential conductance maps in different magnetic fields recorded along the black arrow in Fig. 2A. Obviously, the LLs developed and the features of the quasibound states became obscure with increasing the magnetic fields. The spatial confinement of the angled wedge has two important impacts on the Landau quantization. First, the LLs are shifted away from the charge neutrality point when approaching the summit of the wedge, which is a clear evidence of the LLs bending near the edges (*35-37*). Such a result indicates that the physics of LLs bending still holds even though the boundary of the wedge is not a real edge. Second, the peaks of the LLs become weak and even disappear completely when approaching the vertex of the wedge. The LLs with higher indices disappear at larger distances, as shown in Fig. 3A-3D. This phenomenon arises from the fact that the generation of Landau quantization need quantized Landau orbit length $\sqrt{N}l_B$ and the wave functions of the LLs have their spatial extent $\sim\sqrt{N}l_B$ (here $N$ is index of the LLs and $l_B = \sqrt{\hbar/eB}$ is the magnetic length) (*35-37*). The angled graphene wedge can be divided into three regions, as schematically shown in Fig. 3E. In region I, the distance from the measured position to the boundary of the wedge $l$ is larger than $\sqrt{N}l_B$, then the $N^{th}$ LL is not affected by the spatial confinement. In region II, $l$ is smaller than $\sqrt{N}l_B$, then the $N^{th}$ LL is strongly affected by the spatial confinement and becomes weak with approaching the boundary of the wedge. In region III, we cannot detect the $N^{th}$ LL in the spectra. The spatial extent of the LLs in graphene monolayer was only qualitatively studied in previous studies (*36,37*). The special structure of the angled wedge enables us to quantitatively measure the spatial extent of the LLs $x$ in experiment for the first time ($x$ is obtained by measuring the position where the LLs disappeared in the spectra). In Fig. 3F, we plotted the measured $x$ as a function of $\sqrt{N}l_B$, which agrees quite well with the theoretical result $x = \sqrt{N}l_B$.

The magnetic fields generate cyclotron bending of quasiparticles and affect the quasibound states induced by the spatial confinement in several aspects. Figure 4A , B and C show differential tunneling conductance maps versus magnetic fields measured at $l$ = 25 nm, 27 nm, and 35 nm , respectively. The quasibound states are labeled by $N_1$, $N_2$, $N_3$ and $N_4$, the LLs are labeled by $LL_0$ and $LL_1$. One noticeable feature in Fig. 4A, B and C is that the energy of the quasibound states slightly shifted away from the original value with increasing the magnetic fields. Similar result has been observed previously in transport measurements across two parallel *p*-*n* junctions of graphene (*2*). Such a behavior can be understood quite well with considering the geometric phase induced by magnetic fields on wave functions of the quasibound states (see Fig. S8 for further discussion (*27*)) (*2,13,23-25,38*).

The other obvious phenomenon observed in Fig. 4A, B and C is the sudden and large increase in energy of the quasibound states when a critical magnetic field $B_C$ is achieved, as shown in the black dotted squares. The jump occurs at different magnetic fields in the differential tunneling conductance maps measured at different distance from the summit of the graphene wedge. The increase in energy of the quasibound states at $B_C$ is about one-half the energy spacing between the quasibound states. Such a phenomenon reminds us the half-period shift in the Fabry-Pérot fringe pattern induced by magnetic fields in two parallel *p*-*n* boundaries of graphene (*2,23*). The applied magnetic field bends the trajectories of the charge carriers and changes their incident angle at each *p*-*n* boundary, as shown in Fig. 4D. Because of the Klein tunneling in graphene (i.e., the perfect transmission at normal incidence), the reflection amplitude exhibits a jump in phase by $\pi$ as the sign of the incident angle changes (*2,23*). At zero magnetic field, the contributions of the two *p*-*n* boundaries to the interference cancel with each other because of that the incidence angles at the two interfaces have opposite signs. However, they become imbalanced and the signs of the incidence angles at the two interfaces can be made equal at a finite magnetic field, as shown in Fig. 4D. This effect also can be understood in terms of the Berry phase (*2,13*). The closed momentum space trajectories at $B > B_C$ enclose the Dirac point, which adds a Berry phase of $\pi$ to them, whereas those at $B < B_C$ do not (Fig. 4E). Such an effect affects the properties of the quasibound states and result in the sudden and large increase in energy of the quasibound states observed at $B = B_C$. The value of $B_C$, which depends sensitively on the profile of the *p*-*n*-*p* junction, can be estimated theoretically (*23*). In our experiment, the width of the the confining potential profile at different *l* varies, therefore, we obtained different $B_C$ at different *l*, as shown in Fig. 4A-C. Figure 4F summarizes the measured $B_C$ at different *l* in our experiment. The theoretical $B_C$, which is calculated according to Ref. (*23*) by taking into account the confining potential profile at different *l*, is also plotted, in agreement with experimental data. Even though the theoretical result is obtained in two parallel *p*-*n* boundaries of graphene, it was also demonstrated that the result is still valid in the presence of large-scale spatial fluctuations (*23*). In our experiment, there is an angle ~34 °between the two linear graphene *p*-*n* boundaries. For most of the case, the existence of the angle does not alter the relative sign of incidence angles, as shown in Fig. 4D. Therefore, the main features of our experimental result can be described well by the theory developed in Ref. (*23*). Our experiment demonstrated that the angled wedge is a unique system with the critical magnetic fields for the **π** Berry phase jump depending on distance from summit of the wedge.

Moreover, our experiment also demonstrated that the Full Width at Half Maximum (FWHM) $\delta\varepsilon$ of the quasibound states decreases with increasing the magnetic fields, indicating the increase of the lifetime of quasiparticles $\tau$ ($\tau = \hbar/\delta\varepsilon$), as shown in Fig. S9 (*27*). The lifetime $\tau$ of the quasiparticles increases twice at 12 T comparing to that at 0 T. Such a result indicates that the magnetic field can increase the efficiency of the resonator in trapping the massless Dirac fermions.

In conclusion, we studied spatial confinement, magnetic localization, and their interactions on massless Dirac fermions in the angled graphene wedge. The combination of the spatial and magnetic confinement provides us unprecedented opportunities to explore the effects of spatial confinement on magnetic localization and also the influences of magnetic fields on quasibound states induced by spatial confinement.

**Note added**: During the preparation of the manuscript, we became aware of the work of Christopher Gutierrez, et al (39), which also studied interactions between magnetic fields and spatial confinement on massless Dirac fermions, but in a different system: graphene quantum dot.

## METHODS

**CVD preparation of graphene.** The 25-μm Cu foil was purchased from Alfa Aesar. At first, Cu foil was first electropolished at 1.5 V DC voltage for 60 min, using a mixture of phosphoric acid and ethylene glycol (volume ratio = 3:1) as the electrolyte. Then, we fold the Cu foil to form Cu pocket. Last, the pre-treated Cu foil was loaded into a 2-inch quartz tube of low-pressure chemical vapor deposition furnace for sample growth. The Cu pocket was first heated from room temperature to 1,035 ℃ in 30 min and kept for another 30 min, with 100 sccm (standard cubic centimeter per minute) $H_2$ carrier gas. In the second step, the furnace temperature was set to 1035 ℃ and then the $CH_4$ gas (5 sccm) was introduced into the furnace as carbon source to grow graphene for 10 min. After all growth, the sample was cooled down to room temperature slowly (~20 ℃/min).

**STM and STS measurements.** STM/STS characterizations were performed in ultrahigh vacuum scanning probe microscopes (USM-1300) from UNISOKU. The STM tips were obtained by chemical etching from a wire of Pt/Ir (80/20%) alloys. Lateral dimensions observed in the STM images were calibrated using a standard graphene lattice and a Si (111)-(7×7) lattice and Ag (111) surface. The dI/dV measurements were taken with a standard lock-in technique by turning off the feedback circuit and using a 793-Hz 5mV A.C. modulation of the sample voltage.

**Acknowledgements**

This work was supported by the National Natural Science Foundation of China (Grant Nos. 11674029, 11422430, 11374035), the National Basic Research Program of China (Grants Nos. 2014CB920903, 2013CBA01603), the program for New Century Excellent Talents in University of the Ministry of Education of China (Grant No. NCET-13-0054). L.H. also acknowledges support from the National Program for Support of Top-notch Young Professionals, support from “the Fundamental Research Funds for the Central Universities”, and support from “Chang Jiang Scholars Program”.

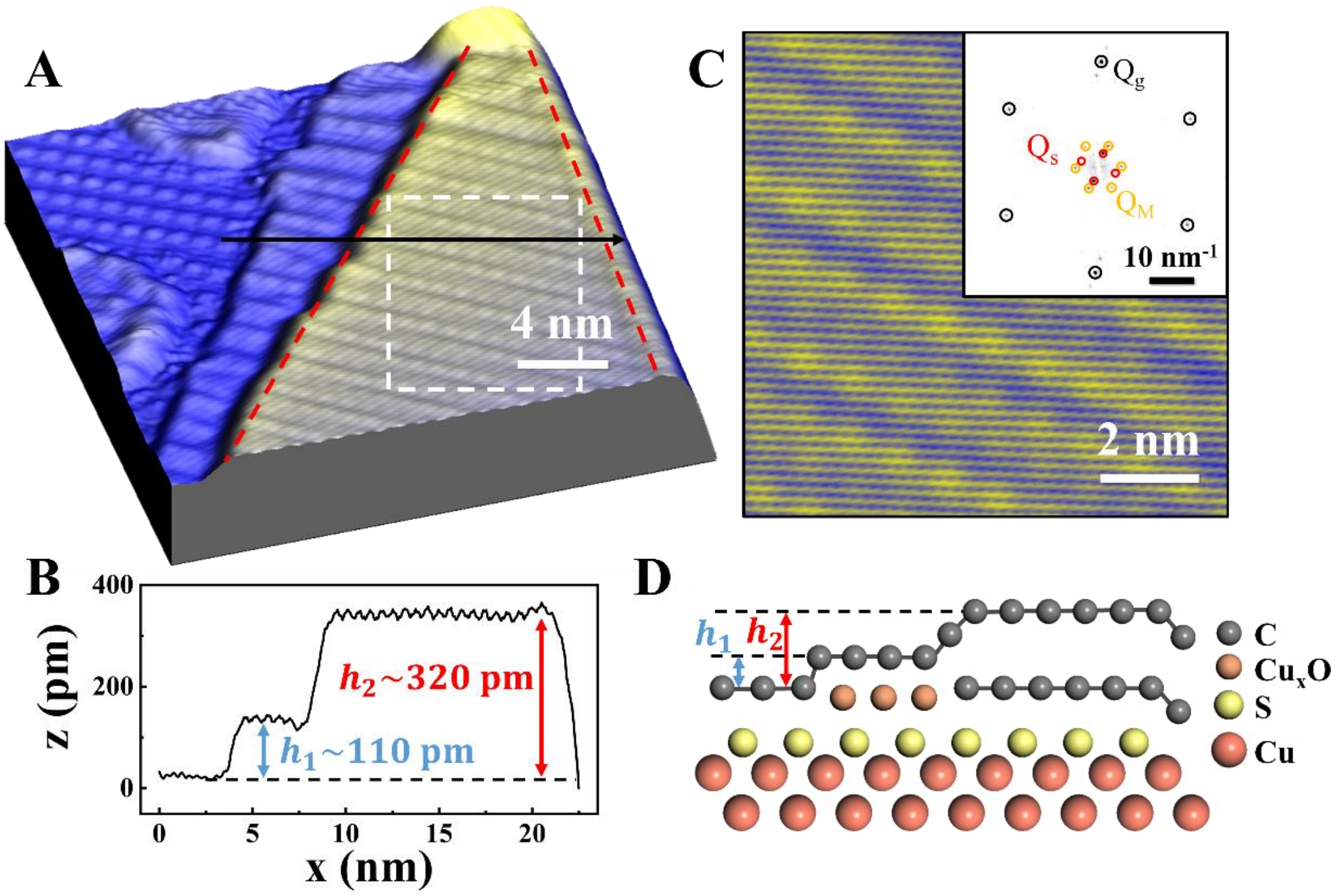

**Figure 1. (A)** A typical three-dimensional STM image of the angled graphene wedge ($V_s = 400\,mV,\ I = 300\,pA$). **(B)** The profile line along the black arrow in (A). **(C),** Zoom-in STM image of white dash square in panel (A) ($V_s = 400\,mV,\ I = 250\,pA$). Inset: FFT image of (C) shows graphene Bragg peaks (black circles), moiré pattern peaks (yellow circles) and sulfur nanocluster superlattices peaks (red circles) **(D),** The schematical side view of the angled graphene wedge.

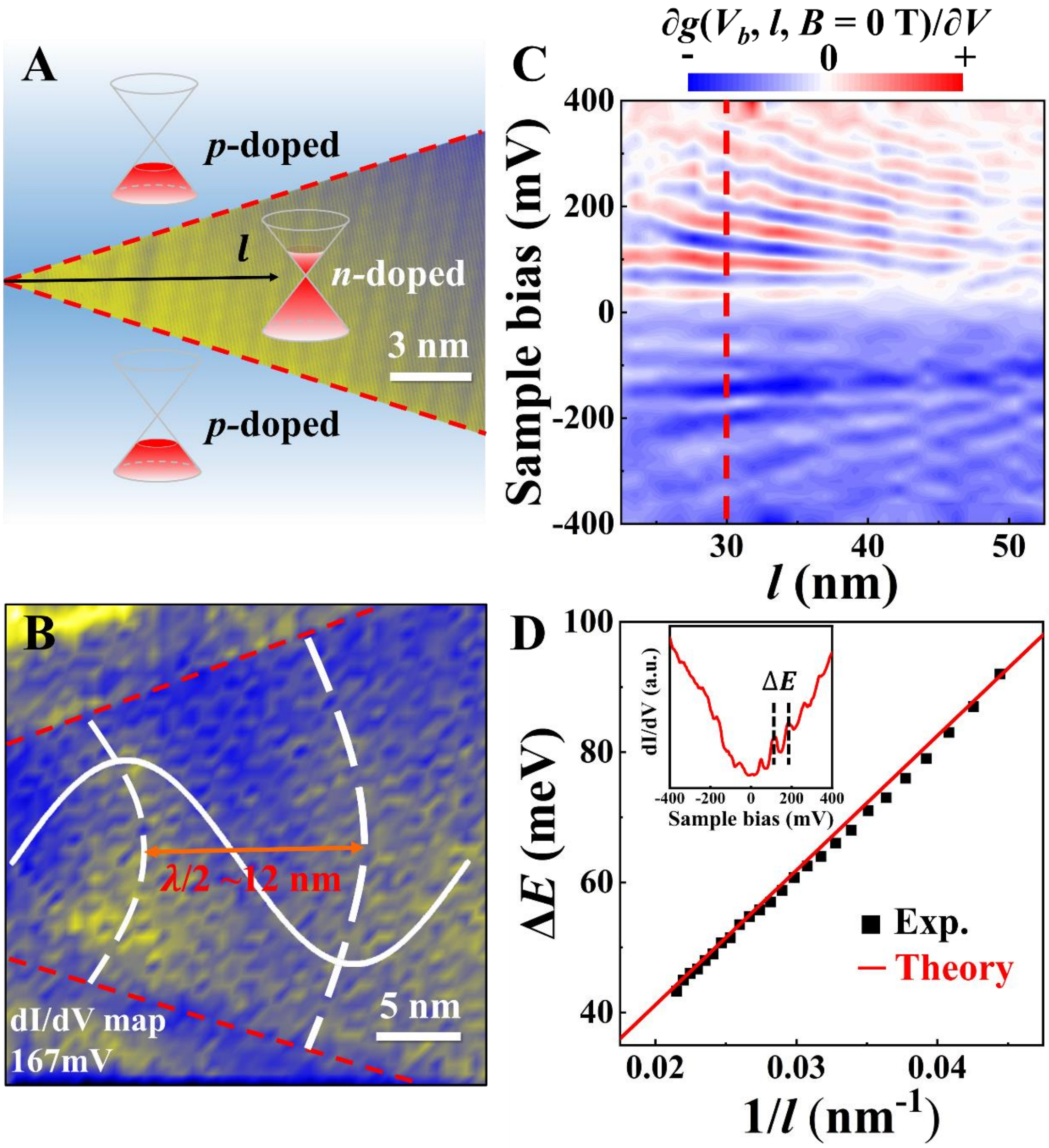

**Figure 2. (A)** Schematic diagram of the *p-n-p* junction. **(B)** A typical STS map recorded at 167 mV, illustrating the spatial distribution of confined Dirac fermions. The wavelength $\lambda$ of the interference pattern is difined in this panel. **(C)** Differential conductance map versus the spatial position *l* obtained along the black arrow in panel (A). **(D)** Energy spacing as a function of the *l*. The red line shows the theoretical result. Inset: a typical STS spectrum measured at $l = 30\ nm$ (red dashed line in panel (C)).

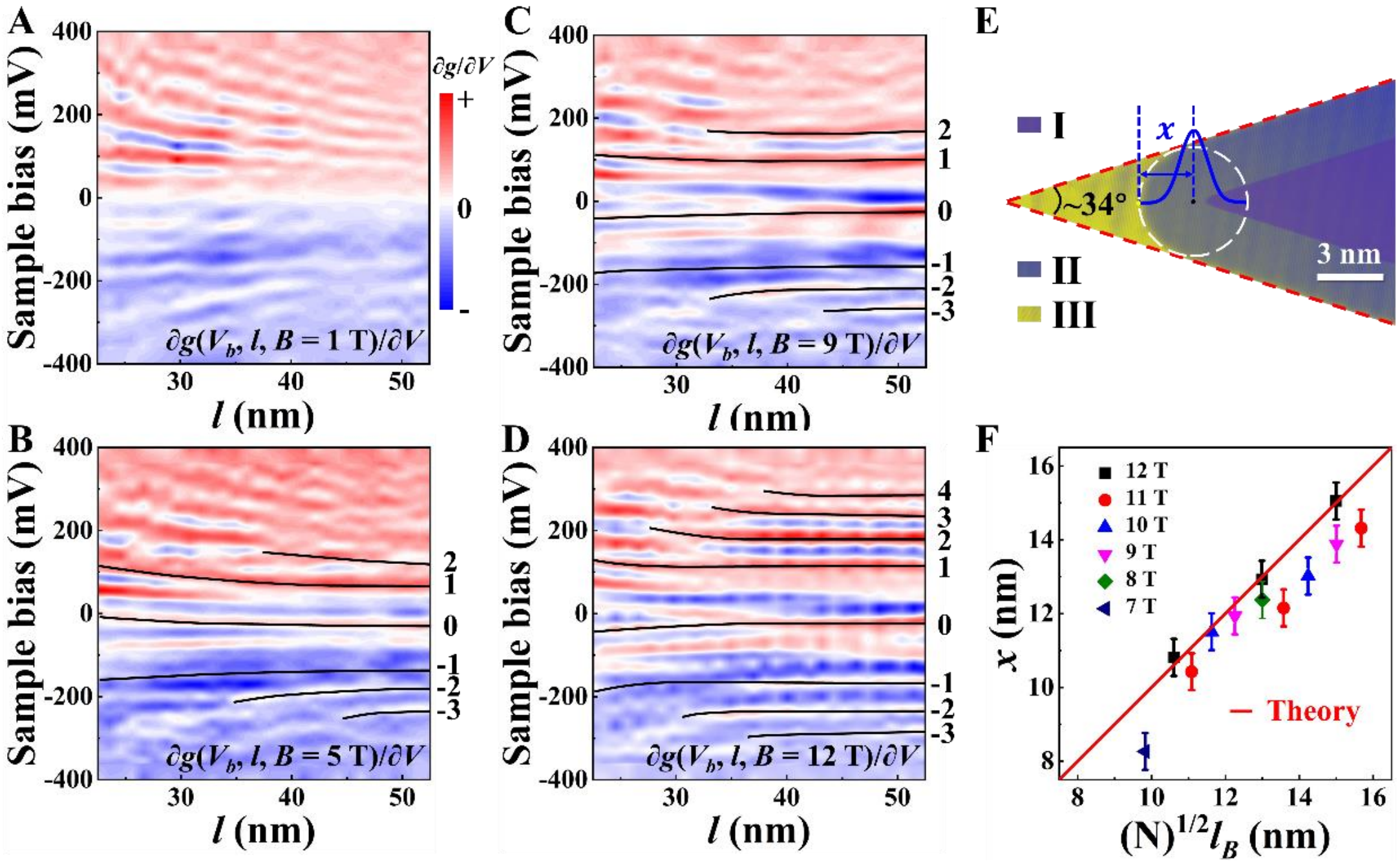

**Figure 3.** Effects of spatial confinement on Landau quantization. **(A to D)** Differential conductance maps versus spatial position *l* obtained along the black arrow in Fig. 2A in different magnetic fields. **(E)** Schematic illustrations of three different regions marked by different colors. The spatial extent length *x* of the $N^{th}$ LL is defined. **(F)** Measured *x* as a function of $\sqrt{N}l_B$, in good agreement with theory (red line).

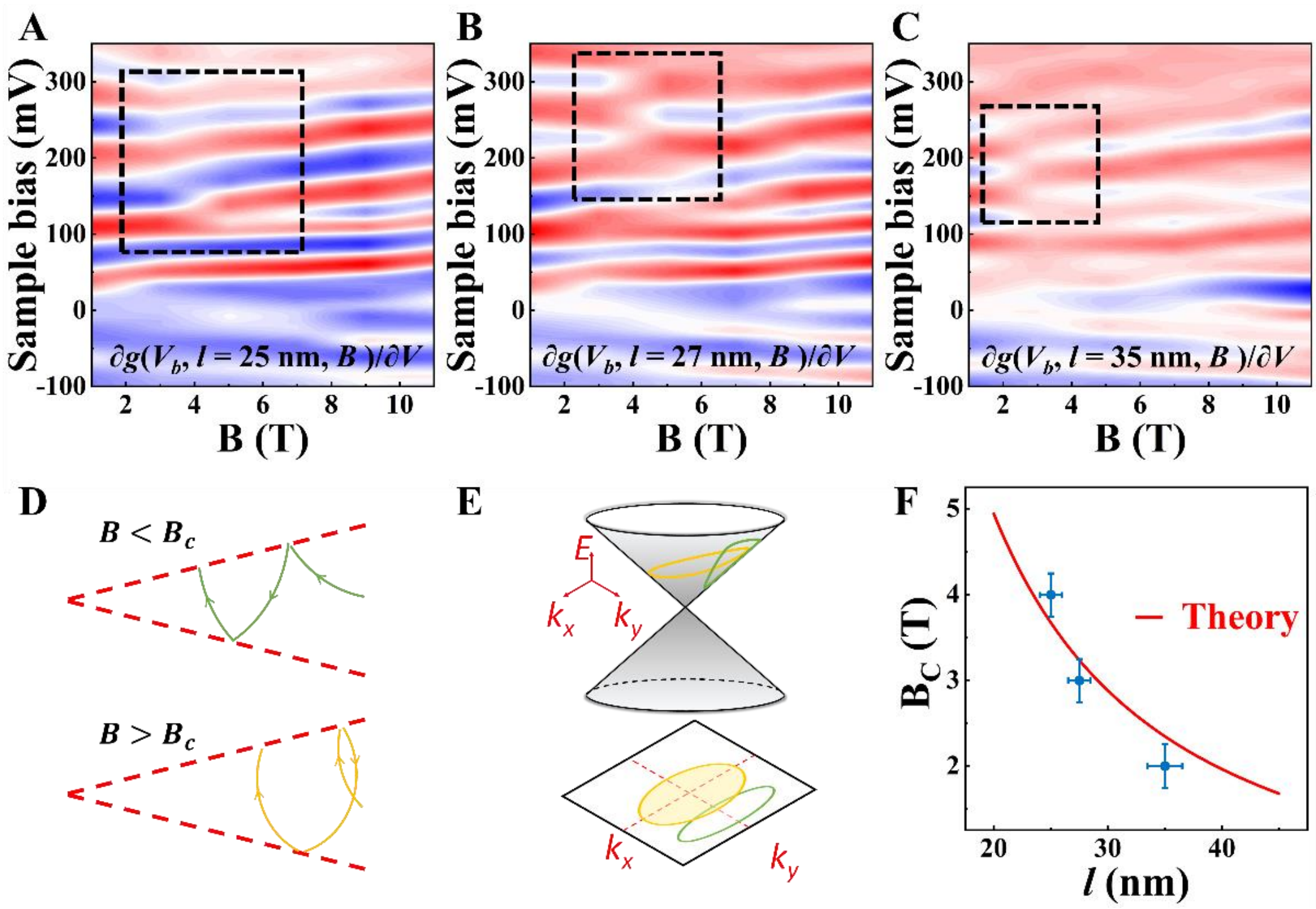

**Figure 4.** Effects of magnetic fields on the quasibound states. **(A to C)** Differential conductance maps versus magnetic field $B$ measured at $l$ = 25 nm,27 nm and 35 nm, respectively. In the black dash squares, there is sudden and large increase in energy of the quasibound states. **(D)** Schematic diagram of charge trajectories in the angled graphene wedge in different applied magnetic fields. Magnetic field will bend trajectories and the incidence angles at the two boundaries will have the same sign when $B > B_C$. **(E)** Schematic diagram of the charge trajectories in momentum space. **(F)** Measured critical magnetic field $B_C$ as a function of the $l$. The theoretical result is also plotted.